\documentclass[12pt,a4paper]{article}

\usepackage{amsmath, amssymb, amsthm}
\usepackage{geometry}
\usepackage{hyperref}
\usepackage{cite}
\usepackage{bm}
\usepackage{bbm}
\usepackage{physics}
\usepackage{enumitem}
\usepackage{titlesec}
\usepackage{abstract}

\usepackage{float}

\hypersetup{
    colorlinks=true,
    linkcolor=blue,
    citecolor=blue,
    urlcolor=blue
}

\newcommand{\vw}{\vec{w}}
\newcommand{\vm}{\vec{m}}
\newcommand{\vn}{\vec{n}}
\newcommand{\veca}{\vec{a}}
\newcommand{\vk}{\vec{k}}

\newcommand{\veta}{\vec{\eta}}
\newcommand{\vlambda}{\vec{\lambda}}

\newcommand{\vS}{\vec{S}}
\newcommand{\vH}{\vec{H}}
\newcommand{\vX}{\vec{X}}

\newcommand{\vLambda}{\vec{\Lambda}}

\title{\textbf{Free-Field Construction of Heterotic String\\
Compactified on Calabi--Yau Orbifolds via \\
Correspondence with $\mathcal{N}{=}2$ SCFT Minimal Models}}

\author{
  Alexander Belavin$^{1}$, Doron Gepner$^{2}$\quad  and Grigory Makarov$^{3}$\\
  \small $^{1}$Landau Institute for Theoretical Physics, 142432, Chernogolovka, Russia\\
  \small $^{2}$Department of Particle Physics and Astrophysics,\\
  \small Weizmann Institute, Rehovot 76100, Israel\\
  \small $^{3}$Skolkovo Institute of Science and Technology, 121205, Moscow, Russia
}

\date{}

\begin{document}
\maketitle

\begin{abstract}
We establish a correspondence between the free-field construction and the minimal-model
construction of the Calabi--Yau sector of the four-dimensional heterotic string compactified
on Berglund--H\"{u}bsch type Calabi--Yau manifolds and their orbifolds. For Fermat-type
polynomials the Calabi--Yau vertex operators expressed in terms of free fields are shown to
correspond to products of primary fields of $\mathcal{N}{=}2$ minimal models. Using this
correspondence we verify modular invariance of the free-field construction and extend it to
Berglund--H\"{u}bsch Calabi--Yau orbifolds, deriving the conditions on complete vertex
operators that parallel those of the minimal-model construction.
\end{abstract}

\tableofcontents

\section{Introduction}

Heterotic string theory is a hybrid of the $\mathcal{N}{=}1$ fermionic string and the bosonic
string. The construction of the ten-dimensional heterotic string was first proposed in
\cite{Candelas1985, Gross1985}. The resulting theory possesses $\mathcal{N}{=}1$ supersymmetry and gauge
symmetry $E(8)\times E(8)$ or $SO(32)$.

To obtain a four-dimensional theory the extra six dimensions must be compactified on
Calabi--Yau manifolds; such a compactification preserves $\mathcal{N}{=}1$ supersymmetry.
In \cite{Gepner1987a,Gepner1987b,Gepner1988} (see also \cite{Blumenhagen2013}) it was conjectured that compactification on a
Calabi--Yau manifold corresponds to compactification on an $\mathcal{N}{=}2$ SCFT with
central charge $c=9$.  The correspondence between the geometrical properties of the Calabi–Yau manifold, such as its cohomology, and the massless spectrum of the theory is deeply rooted in the structure of the $\mathcal{N}=2$ chiral rings and spectral flow \cite{Lerche1989}. The resulting four-dimensional theory possesses $\mathcal{N}{=}1$
supersymmetry and gauge symmetry $E(8)\times E(6)$.

For the special subclass of Berglund--H\"{u}bsch Calabi--Yau manifolds (or their orbifolds)
this compactification can be realised as a product of five $\mathcal{N}{=}2$ SCFT minimal
models  \cite{Gepner1987a,Gepner1987b,Gepner1988, Gepner1989}. The precise geometric interpretation of such products as Calabi-Yau hypersurfaces in weighted projective spaces was revealed via the Landau-Ginzburg renormalization group flows in \cite{Greene1989}. In this construction modular invariance is verified explicitly using the characters of
the $\mathcal{N}{=}2$ minimal models.

In \cite{Belavin2025} a construction was proposed in which the compact sector is realised by
a free-field representation of the $\mathcal{N}{=}2$ SCFT. This construction is applicable to
any Berglund--H\"{u}bsch Calabi--Yau manifold \cite{Berglund1993, Kreuzer2000, Krawitz2009}. It employs the Batyrev--Borisov
combinatorial data, namely dual lattices and Batyrev polytopes, whose vectors appear
explicitly in the construction of physical vertex operators.

In the present paper we discuss the correspondence between the two constructions---the
minimal-model construction and the free-field construction. The Calabi--Yau vertex operators
in terms of free fields turn out, in the Fermat case, to correspond to products of primary
fields.

This observation, first of all, allows us to verify that the free-field construction satisfies the
condition of modular invariance. Moreover, building on the discovered correspondence, we
generalise the construction to the case of Berglund--H\"{u}bsch Calabi--Yau orbifolds in
such a way that the modular-invariance conditions from the minimal-model construction are
satisfied.

A Berglund--H\"{u}bsch Calabi--Yau orbifold is specified by the choice of an admissible
group $G$. From it we find the Batyrev--Borisov combinatorial data for orbifolds. The dual
lattices will contain vectors corresponding to elements of the group $G$ and elements of the
mirror group $G^*$, and the dual polytopes will include points corresponding to deformations
and mirror deformations.

Furthermore, we find that upon compactification on a Calabi--Yau orbifold, the left and right vertex algebras of the heterotic string differ significantly: the left algebra is constructed based on the combinatorial data for the minimal admissible group, whereas the right vertex operators include vectors belonging to the lattices constructed from the admissible group $G$ that defines the orbifold.

Thus, it turns out that the heterotic string compactified on a Calabi--Yau orbifold is not only a hybrid of the fermionic and bosonic strings, but also a hybrid of two Calabi--Yau manifolds: the original Calabi--Yau manifold as a hypersurface in a weighted projective space in the left (holomorphic) sector, and its orbifold in the right sector.

Finally, having defined the conditions on the complete vertex operators as products of their left and right parts, we confirm the matching of the massless spectra in the orbifold case, thanks to the discovered correspondence between the free-field construction and the minimal-model construction. We also confirm that our proposed conditions on the complete vertex operators correspond to the modular invariance conditions from the minimal-model construction, which validates this property of the free-field construction for Calabi--Yau manifolds defined by a Fermat-type polynomial.

\section{Free-Field Construction of the 4D Heterotic String}
\label{sec:Gmin}
In this section we briefly review the free-field construction of the four-dimensional heterotic
string compactified on Berglund--H\"{u}bsch Calabi--Yau manifolds \cite{Berglund1993, Krawitz2009}. A detailed exposition
can be found in \cite{Belavin2025}. We discuss the definition of Berglund--H\"{u}bsch
Calabi--Yau manifolds and the procedure for obtaining the Batyrev--Borisov combinatorial
data.

\subsection{Berglund--H\"{u}bsch Calabi--Yau Manifolds}

Berglund--H\"{u}bsch Calabi--Yau manifolds are defined as hypersurfaces in a weighted
projective space
\begin{equation}
  \mathbb{P}^4_{\vk} = \bigl\{y\in\mathbb{C}^5\;\big|\;
    (\lambda^{k_1}y_1,\ldots,\lambda^{k_5}y_5)\sim(y_1,\ldots,y_5)\;\forall\,\lambda\in\mathbb{C}\bigr\}.
\end{equation}
The hypersurface is defined by the equation
\begin{equation}
  W_0(y)=\sum_{i=1}^{5}\prod_{j=1}^{5}y_j^{A_{ij}}=0,
\end{equation}
where $A_{ij}$ is a non-degenerate matrix and the polynomial $W_0(y)$ is quasi-homogeneous.
The latter condition means that $\sum_j A_{ij}k_j=d$, where $d=\sum k_i$ for Calabi--Yau
manifolds due to the vanishing of the first Chern class.

The case of most interest in this paper is the Fermat-type polynomial
$W_0(y)=\sum_i y_i^{A_{ii}}$, i.e.\ $A_{ij}=A_{ii}\delta_{ij}$, since such a Calabi--Yau
corresponds to a product of $\mathcal{N}{=}2$ SCFT minimal models.

For a Fermat-type polynomial $W_0$ the maximal set of deformations $\vec{m}$ is defined by
\begin{equation}\label{eq:defcond}
  \sum_i k_i m_i = d,\qquad 0\le m_i\le A_{ii}-2,\quad m_i\in\mathbb{Z},
\end{equation}
where each vector $\vec{m}$ corresponds to a deformation of $W_0$ by the monomial
$\prod_i y_i^{m_i}$, giving a family of Calabi--Yau manifolds.

One then chooses an admissible group $G$, which is a subgroup of the maximal admissible group
\begin{equation}\label{eq:Gmaxadm}
  G^{\max}_{\mathrm{adm}}=\left\{
    \left.\left(\exp\!\left(2\pi i\sum_{j=1}^5 B_{1j}w_j\right),\ldots,
                 \exp\!\left(2\pi i\sum_{j=1}^5 B_{5j}w_j\right)\right)\,\right|\,
    w_i\in\mathbb{Z},\;
    \sum_{i,j}B_{ij}w_j\in\mathbb{Z}
  \right\},
\end{equation}
where $B_{ij}=(A^{-1})_{ij}$. Its elements act on the coordinates of the weighted projective
space as
\begin{equation}
  y_i\;\longrightarrow\; y_i\exp\!\left(\sum_{j=1}^5 B_{ij}w_j\right),
\end{equation}
which for a Fermat polynomial reduces to
\begin{equation}
  y_i\;\longrightarrow\; y_i\exp\!\left(\frac{2\pi i\,w_i}{A_{ii}}\right).
\end{equation}

For brevity we shall denote group elements by the corresponding vectors $\vec{w}$. The group
\eqref{eq:Gmaxadm} contains the minimal group $G^{\min}_{\mathrm{adm}}$, generated by
$\vec{w}=(1,1,1,1,1)$, as a subgroup; it corresponds to the symmetry of the weighted
projective space $y_i\to y_i\exp(2\pi i k_i/d)$.

The choice of admissible group $G^{\min}_{\mathrm{adm}}\subseteq G\subseteq G^{\max}_{\mathrm{adm}}$
imposes additional conditions on the set of admissible deformations:
\begin{equation}
  \sum_{i,j=1}^5 m_i B_{ij} w_j \in\mathbb{Z}\qquad\forall\,\vec{w}\in G.
\end{equation}
The group $G$ has a set of generators $\vec{\gamma}_a$ such that every element decomposes as
$\vec{w}=\sum_a t^a\vec{\gamma}_a$.

The mirror Calabi--Yau, defined in $\mathbb{P}^4_{\vk^*}$, is given by the mirror polynomial
\begin{equation}
  W^*_0(y)=\sum_{i=1}^5\prod_{j=1}^5 y_j^{(A^T)_{ij}}=0.
\end{equation}
In the Fermat case $A^T=A$ and $\vk^*=\vk$.

According to Krawitz \cite{Krawitz2009}, the generators of the mirror group $G^*$ can be defined via the
deformations $\vec{m}$ of the original Calabi--Yau as
\begin{equation}
  \left(\exp\!\left(2\pi i\sum_j B_{1j}m_j\right),\ldots,
         \exp\!\left(2\pi i\sum_j B_{5j}m_j\right)\right)\in G^*.
\end{equation}
Consequently, mirror deformations $\vec{n}$ must satisfy
\begin{equation}
  \sum_{i,j=1}^5 m_i B_{ij} n_j \in\mathbb{Z}
\end{equation}
for every deformation $\vec{m}$ of the original Calabi--Yau.

 A general algorithm for systematically finding all such admissible groups $G$ and their mirrors $G^*$, along with the complete sets of corresponding original and mirror deformations to construct all possible mirror pairs of Berglund–Hübsch orbifolds, was developed in \cite{Aleshin2026}.
\subsection{Combinatorial Data for $G^{\min}_{\mathrm{adm}}$}

For the case of the minimal admissible group $G=G^{\min}_{\mathrm{adm}}$ one obtains the
Batyrev--Borisov combinatorial data (see \cite{Belavin2025,Borisov2004, Borisov2011}) from the given polynomial $W_0$. The generalisation to
Calabi--Yau orbifolds will be discussed in Section~\ref{sec:orbifolds}.

First, one introduces five-dimensional integer lattices $M_0$ and $N_0$ with basis vectors
$\vec{u}_i$ and $\vec{v}_j$ satisfying
\begin{equation}
  \vec{u}_i\cdot\vec{v}_j = A_{ij}.
\end{equation}
We choose basis vectors as
\begin{equation}\label{eq:basis}
  (\vec{u}_i)_j = A_{ij},\qquad (\vec{v}_i)_j = \delta_{ij}.
\end{equation}
We also define vectors $\vec{a}^+$ and $\vec{a}^-$ as
\begin{equation}
  \vec{a}^+ = \frac{1}{d^*}\sum_i k^*_i\,\vec{u}_i,\qquad
  \vec{a}^- = \frac{1}{d}\sum_j k_j\,\vec{v}_j.
\end{equation}
Note that $\vec{u}_i\cdot\vec{a}^- = \vec{a}^+\cdot\vec{v}_j = \vec{a}^+\cdot\vec{a}^- = 1$.
For the chosen bases \eqref{eq:basis} we have $a^+_i=1$ and $a^-_i=k_i/d$.

The dual cones are defined as
\begin{equation}
  K = \sum_i\mathbb{Q}_{\ge 0}\,\vec{u}_i,\qquad
  K^* = \sum_j\mathbb{Q}_{\ge 0}\,\vec{v}_j.
\end{equation}

For the minimal admissible group the lattice $N$ is defined as the minimal extension of $N_0$:
\begin{equation}\label{eq:NGmin}
    N(G^{\min}_{\mathrm{adm}}) = N_0 + \mathbb{Z}\,\vec{a}^-.
\end{equation}
A basis $\vec{e}^*_\beta$, $\beta=1,\ldots,5$, of the lattice $N$ is then found, and the basis
$\vec{e}_\alpha$ of the dual lattice $M(G^{min}_{adm})$ follows from
\begin{equation}
  \vec{e}_\alpha\cdot\vec{e}^*_\beta = \delta_{\alpha\beta}.
\end{equation}
Finally, the Batyrev dual polytopes $\Delta$ and $\Delta^*$ are defined as the set of lattice
points (or dual lattice points) belonging to the cone and satisfying:
\begin{align}
  \Delta  &= \{\vm\in M(G^{min}_{adm}) \mid \vm\in K,\; \vm\cdot\vec{a}^-=1\},\\
  \Delta^* &= \{\vn\in N(G^{min}_{adm}) \mid \vn\in K^*,\;\vec{a}^+\cdot\vn=1\}.
\end{align}
Points $\vm\in\Delta$ satisfying $m_i\le A_{ii}-2$ (in the Fermat case) correspond to
admissible deformations of $W_0$.

\subsection{The Calabi--Yau Sector in the Free-Field Construction}

In the free-field construction the compact Calabi--Yau sector is built as an $\mathcal{N}{=}2$
SCFT with ten free bosonic fields $X^\pm_i(z)$ and ten Majorana fermions $\Psi^\pm_i(z)$
($i=1,\ldots,5$). The latter can be bosonised by introducing five bosonic fields $H_i(z)$ via
$\Psi^\pm_i(z)=\exp(\pm iH_i(z))$. The OPEs of the fields are
\begin{align}
  X^+_i(z)\,X^-_j(0) &= \delta_{ij}\log z+\cdots,\\
  \Psi^+_i(z)\,\Psi^-_j(0) &= \delta_{ij}\,z^{-1}+\cdots,\\
  H_i(z)\,H_j(0) &= -\delta_{ij}\log z+\cdots.
\end{align}
The currents of the $\mathcal{N}{=}2$ super-Virasoro algebra are expressed in terms of these
fields as
\begin{align}
  T_{\mathrm{CY}}(z) &= \sum_{i=1}^5\Bigl[
    \partial X^+_i\,\partial X^-_i
    -\tfrac{1}{2}(\partial H_i)^2
    -\tfrac{1}{2}\bigl(a^+_i\partial^2 X^-_i + a^-_i\partial^2 X^+_i\bigr)
  \Bigr],\\
  G^+_{\mathrm{CY}}(z) &= \sqrt{2}\sum_{i=1}^5\bigl[
    \Psi^+_i\,\partial X^-_i - \partial\Psi^+_i\,a^-_i\bigr],\\
  G^-_{\mathrm{CY}}(z) &= \sqrt{2}\sum_{i=1}^5\bigl[
    \Psi^-_i\,\partial X^+_i - \partial\Psi^-_i\,a^+_i\bigr],\\
  J_{\mathrm{CY}}(z) &= \partial\!\left[\sum_{i=1}^5\bigl(
    iH_i + a^-_i\partial X^+_i - a^+_i\partial X^-_i\bigr)\right]
    = \partial H_{\mathrm{CY}}(z).
\end{align}
The central charge of the Calabi--Yau sector in this construction is
\begin{equation}\label{eq:cCY}
  c_{\mathrm{CY}} = 15 - 6\sum_{i=1}^5 a^+_i a^-_i = 9.
\end{equation}
Physical states carry an additional condition: the corresponding vertex operators must be
cohomologies of the Borisov differential \cite{Belavin2025,Borisov2004, Borisov2011}
\begin{equation}\label{eq:D}
  D = \sum_{i=1}^5 D_{\vec{u}_i} + \sum_{j=1}^5 D_{\vec{v}_j},
\end{equation}
where
\begin{equation}
  D_{\vec{u}_i} = \oint dz\;\vec{u}_i\cdot\vec{\Psi}^-(z)\,
    \exp(\vec{u}_i\cdot\vec{X}^-(z)),\qquad
  D_{\vec{v}_j} = \oint dz\;\vec{v}_j\cdot\vec{\Psi}^+(z)\,
    \exp(\vec{v}_j\cdot\vec{X}^+(z)).
\end{equation}
The differential \eqref{eq:D} commutes with the $\mathcal{N}{=}2$ super-Virasoro currents.

A useful fact is that the exponential factor
\begin{equation}
  \exp\!\bigl(\vm\cdot\vX^-(z)+\vn\cdot\vX^+(z)+i\vS\cdot\vH(z)\bigr)
\end{equation}
has conformal dimension and $U(1)$ charge
\begin{align}
  \Delta_{\mathrm{CY}} &= \vm\cdot\vn + \tfrac{1}{2}(\vm\cdot\vec{a}^- + \vn\cdot\vec{a}^+)
    + \frac{\vS^2}{2}, \label{eq:DeltaCY}\\
  Q_{\mathrm{CY}} &= \vm\cdot\vec{a}^- - \vn\cdot\vec{a}^+ + \sum_i S_i, \label{eq:QCY}
\end{align}
where $\vm\in M(G^{min}_{adm})$, $\vn\in N(G^{min}_{adm})$, $S_i=0,\pm1$ in the NS sector, and
$\vm\in M(G^{min}_{adm})\pm\tfrac{1}{2}\vec{a}^+$, $\vn\in N(G^{min}_{adm})\pm\tfrac{1}{2}\vec{a}^-$, $S_i=\pm\tfrac{1}{2}$ in
the R sector.

In the present paper we establish a correspondence between such vertex operators and
products of primary fields in the minimal-model approach.

\subsection{Vertex Algebra for the Minimal Admissible Group}

The left (holomorphic) sector of the four-dimensional heterotic string is the product of a
four-dimensional $\mathcal{N}{=}1$ SCFT (the spacetime subsector, $c^L_{\mathrm{ST}}=6$)
and the Calabi--Yau subsector ($c^L_{\mathrm{CY}}=9$), giving the critical central charge
$c=15$.

The spacetime subsector contains four bosonic fields $x^\mu(z)$ and four Majorana fermions
$\psi^\mu(z)$ ($\mu=0,\ldots,3$), which can be bosonised by introducing two bosonic fields
$\mathcal{H}_a(z)$:
\begin{equation}
  \frac{1}{\sqrt{2}}\bigl(\pm\psi^0(z)+\psi^1(z)\bigr) = e^{\pm i\mathcal{H}_1(z)},\qquad
  \frac{1}{\sqrt{2}}\bigl(\psi^2(z)\pm i\psi^3(z)\bigr) = e^{\pm i\mathcal{H}_2(z)}.
\end{equation}
A general left vertex operator takes the form
\begin{equation}\label{eq:VL}
  V^L_{q,\vlambda,\vm,\vn,\vS}(z)
  = P(\partial^k x^\mu,\partial^l \mathcal{H}_a,\partial^r H_i,\partial^s X^+_i,\partial^t X^-_i)\,
    e^{q\phi(z)}\,
    e^{i\vlambda\cdot\vec{\mathcal{H}}(z)+\vm\cdot\vX^-+\vn\cdot\vX^++i\vS\cdot\vH+ip_\mu x^\mu},
\end{equation}
with conformal dimension
\begin{equation}
  \Delta^L = -\frac{q(q+2)}{2} + \frac{\vlambda^2}{2} + \frac{(p^\mu)^2}{2}
    + \Delta^L_{\mathrm{CY}} + N^L = 1.
\end{equation}
Here $q$ specifies the picture; we work in canonical pictures: $q=-1$ for NS states and
$q=-1/2$ for R states. Also, $\vlambda$ is a vector belonging to one of the conjugacy classes
of $SO(1,3)$: $[0]$, $[V]$ for NS states or $[S]$, $[C]$ for R states.

The vertex operators \eqref{eq:VL} are subject to the following conditions.  First, the total left-moving $\mathcal{N}{=}1$ super-Virasoro algebra is defined as the diagonal subalgebra of the direct sum of the $\mathcal{N}{=}1$ super-Virasoro algebras of the space-time and the compact Calabi--Yau subsectors. Its currents are given by the sum of the corresponding super-Virasoro currents from both subsectors:
\begin{equation}\label{eq:sVirLeft}
    T(z)=T_{ST}(z)+T_{CY}(z); \qquad G(z) = G_{ST}(z)+G_{CY}(z).
\end{equation}
Furthermore, the generators of this algebra --- namely, the modes of the currents \eqref{eq:sVirLeft} --- commute with the Borisov differential \eqref{eq:D}. Since this diagonal algebra acts on the space of left vertex operators, they must belong entirely to either the NS or R representation of the total $\mathcal{N}{=}1$ super-Virasoro algebra.

Second, left vertex operators must be BRST cohomologies and Borisov
cohomologies with respect to the differential \eqref{eq:D}.  Then, among the massless left
vertex operators we find  mutually local vertex operators corresponding to the currents of
$\mathcal{N}{=}1$ super-Poincar\'{e}:
\begin{equation}\label{eq:J}
  J^+_{\vec{\sigma}}(z) = \exp\!\left(-\tfrac{1}{2}\phi + i\vec{\sigma}\cdot\vec{\mathcal{H}}
    + \tfrac{1}{2}H^L_{\mathrm{CY}}\right),\qquad
  J^-_{\dot{\vec{\sigma}}}(z) = \exp\!\left(-\tfrac{1}{2}\phi + i\dot{\vec{\sigma}}\cdot\vec{\mathcal{H}}
    - \tfrac{1}{2}H^L_{\mathrm{CY}}\right),
\end{equation}
where $\vec{\sigma}\in[S]$ and $\dot{\vec{\sigma}}\in[C]$.  The integrals of currents \eqref{eq:J} commute both with BRST and with Borisov differential \eqref{eq:D} and have the same commutation relations as generators of $\mathcal{N}=1$ super-Poincar\'{e}. (see \cite{Belavin2025} for more details)

We require the left vertex operators \eqref{eq:VL} to be mutually local with the currents
\eqref{eq:J} --- the GSO$_L$ condition:
\begin{equation}
  q + \sum_i\lambda_i + Q^L_{\mathrm{CY}} \in 2\mathbb{Z}.
\end{equation}

 As a result of these constraints, the elements of the left vertex algebra belong both to  BRST and Borisov cohomologies as well as to a representation of $\mathcal{N}=1$ super-Poincar\'{e}.

The right (antiholomorphic) sector of the four-dimensional heterotic string contains the
spacetime subsector with four free bosons $x^\mu(\bar z)$. To obtain the critical central
charge $c^R=26$ we add eight free bosonic fields $Y^I(\bar z)$ compactified on the $E(8)$ root
torus, as well as five free bosons $\Phi^\alpha(\bar z)$ compactified on the $SO(10)$ root
torus, and finally an $\mathcal{N}{=}2$ SCFT with $c=9$ in the form of the Calabi--Yau
subsector.

A general right vertex operator takes the form
\begin{equation}\label{eq:VR}
  V^R_{\veta,\vLambda,\vm,\vn,\vS}(\bar z)
  = P(\bar\partial^k\bar x^a,\bar\partial^l Y^I,\bar\partial^{l'}\Phi^\alpha,
       \bar\partial^r H_i,\bar\partial^s\bar X^+_i,\bar\partial^t\bar X^-_i)\,
    e^{i\veta\cdot\vec{Y}+i\vLambda\cdot\vec{\Phi}+\vm\cdot\vX^-+\vn\cdot\vX^++i\vS\cdot\vH+ip_\mu x^\mu},
\end{equation}
where $\veta$ belongs to the $E(8)$ root lattice, $\vLambda$ belongs to the $SO(10)$ weight
lattice, and the conformal dimension is
\begin{equation}
  \Delta^R = \frac{\veta^2}{2} + \frac{\vLambda^2}{2} + \frac{(p^\mu)^2}{2}
    + \Delta^R_{\mathrm{CY}} + N^R = 1.
\end{equation}
Among the right vertex operators \eqref{eq:VR} we retain only those that are BRST
cohomologies and Borisov cohomologies.

Among massless right vertex operators we find vertex operators corresponding to the $E(8)$
algebra currents:
\begin{equation}\label{eq:E8}
  J^I(\bar z) = i\bar\partial Y^I,\qquad
  J_{\vec{\epsilon}} = \exp(i\epsilon_I Y^I);
\end{equation}
to $SO(10)$ currents:
\begin{equation}\label{eq:SO10}
  J^\alpha(\bar z) = i\bar\partial\Phi^\alpha,\qquad
  J_{\vec{\rho}} = \exp(i\rho_\alpha\Phi^\alpha);
\end{equation}
and to the $U(1)$ current:
\begin{equation}\label{eq:U1}
  J_{\mathrm{CY}}(\bar z) = \bar\partial H^R_{\mathrm{CY}}
    = \bar\partial\sum_{i=1}^5\bigl(i\bar H_i + a^-_i\bar X^+_i - a^+_i\bar X^-_i\bigr).
\end{equation}

Thus we find generators of $E(8)\times SO(10)\times U(1)$ in the right sector  as integrals of these currents, which commute with Borisov differential and the BRST charge \cite{Belavin2025}. Moreover,
additional currents allow us to extend the gauge symmetry to $E(8)\times E(6)$:
\begin{equation}\label{eq:SO10spinors}
  J^+_\omega(\bar z) = \exp\!\bigl(i\omega^\alpha\Phi_\alpha + \tfrac{1}{2}H^R_{\mathrm{CY}}\bigr),
  \qquad
  J^-_{\dot\omega}(\bar z) = \exp\!\bigl(i\dot\omega^\alpha\Phi_\alpha - \tfrac{1}{2}H^R_{\mathrm{CY}}\bigr).
\end{equation}
We require the right vertex operators to be mutually local with the gauge currents listed
above --- the GSO$_R$ condition:
\begin{equation}
  \sum_i\Lambda_i + Q^R_{\mathrm{CY}} \in 2\mathbb{Z}.
\end{equation}

Finally, from the sets of left and right vertex operators we construct diagonal (with respect to
the Calabi--Yau part) complete vertex operators:
\begin{equation}
  V(z,\bar z) = V^L_{q,\vlambda,\vm^L,\vn^L,\vS^L}(z)
    \otimes V^R_{\veta,\vLambda,\vm^R,\vn^R,\vS^R}(\bar z),
\end{equation}
where $\vm^{L,R}\in M(G^{min}_{adm})$, $\vn^{L,R}\in N(G^{min}_{adm})$ (or $\vm^{L,R}\in M(G^{min}_{adm})\pm\tfrac{1}{2}\vec{a}^+$,
$\vn^{L,R}\in N(G^{min}_{adm})\pm\tfrac{1}{2}\vec{a}^-$), and diagonality means
\begin{equation}
  \vm^L = \vm^R,\qquad \vn^L = \vn^R.
\end{equation}

We then act on this set with $\mathcal{N}{=}1$ super-Poincar\'{e} generators \eqref{eq:J}
from the left and/or gauge group generators from the right, thereby obtaining the complete set
of physical vertex operators. Their Calabi--Yau parts may now differ by a spectral flow:
\begin{equation}
  \vm^L = \vm^R + \frac{N}{2}\vec{a}^+,\qquad
  \vn^L = \vn^R - \frac{N}{2}\vec{a}^+,\quad N\in\mathbb{Z}.
\end{equation}

This set of complete vertex operators will satisfy the conditions of mutual locality and
modular invariance. Modular invariance of this construction will be verified in the next
section from the correspondence with the minimal-model construction, for which this property
is proven.

\subsection{List of Types of Complete Massless Vertex Operators}
\begin{itemize}[leftmargin=*, label=\textbullet]
\item \textbf{Gravitational supermultiplet} ($[V]\times[0]$ + superpartners):
\begin{equation}
  e^{-\phi(z)}\Psi^\mu(z)\otimes i\bar\partial\bar X^\nu(\bar z).
\end{equation}

\item \textbf{Gauge supermultiplet} --- vector bosons ($[V]\times[0]$ + superpartners):
\begin{equation}
  e^{-\phi(z)}\Psi^\mu(z)\otimes[\text{currents of }E(8)\times E(6)](\bar z).
\end{equation}

\item \textbf{27-supermultiplet} ($[0]\times[V]$ + superpartners and $E(6)$ partners):
\begin{equation}
  e^{-\phi(z)}\exp[\vm\cdot\vX^-](z)\otimes\exp[\vm\cdot\vX^-+i\vLambda\cdot\vec{\Phi}](\bar z),
\end{equation}
where $Q^L_{\mathrm{CY}}=Q^R_{CY}=+1$ (or $-1$ for antiparticles), $\Delta^L_{\mathrm{CY}}=1/2$,
$\Delta^R_{\mathrm{CY}}=1/2$, and $\vm\in\Delta$.

Note that 27-vertex operators with vectors $\vm$ not satisfying $m_i\le A_{ii}-2$ (in the
Fermat case) lie in the image of the Borisov differential and are therefore not physical. Thus
the number of 27-supermultiplet vertex operators equals the number of admissible
deformations of $W_0$, and hence equals the Hodge number $h^{2,1}$.  This direct matching between the primary fields of the $\mathcal{N}=2$ chiral ring and the elements of the Calabi–Yau cohomology ring was fundamentally established in \cite{Lerche1989}.

\item \textbf{$\overline{27}$-supermultiplet} ($[0]\times[V]$ + superpartners and $E(6)$ partners):
\begin{equation}
  e^{-\phi(z)}\exp[\vec{m}\cdot\vX^-](z)\otimes\exp[\vec{n}\cdot\vX^++i\vLambda\cdot\vec{\Phi}](\bar z),
\end{equation}
where $Q^L_{\mathrm{CY}}=+1$, $Q^R_{CY}=-1$ (or $Q^L_{CY}=-1$, $Q^R_{CY}=+1$ for antiparticles), $\Delta^L_{\mathrm{CY}}=1/2$, $\Delta^R_{\mathrm{CY}}=1/2$.

Note that the $\overline{27}$-supermultiplet is unique for Fermat-type Calabi--Yau in the
case of the minimal admissible group; indeed, in that case there is a unique mirror
deformation. Vertex operators of type $(c,c)$ correspond to 27, and those of type $(c,a)$
correspond to $\overline{27}$.

\item \textbf{Gauge singlets} ($[0]\times[0]$ + superpartners):
\begin{align}
  &e^{-\phi(z)}P^L_{\mathrm{CY}}[\ldots](z)
    \exp[\vm^L\cdot\vX^-+\vn^L\cdot\vX^++i\vS^L\cdot\vH](z)\nonumber\\
  &\quad\otimes P^R_{\mathrm{CY}}[\ldots](\bar z)
    \exp[\vm^R\cdot\vX^-+\vn^R\cdot\vX^++i\vS^R\cdot\vH](\bar z),
\end{align}
where $Q^L_{\mathrm{CY}}=+1$ (or $-1$ for antiparticles), $\Delta^L_{\mathrm{CY}}=1/2$,
$Q^R_{\mathrm{CY}}=0$, $\Delta^R_{\mathrm{CY}}=1$; $\vm^{L,R}\in M$, $\vn^{L,R}\in N$.
\end{itemize}

\section{Correspondence Between the Two Constructions}

\subsection{Minimal-Model Construction}
\label{subsec:minmod}
In the Gepner construction the role of the $\mathcal{N}{=}2$ SCFT with central charge 9 is
played by a product of $\mathcal{N}{=}2$ SCFT minimal models. Consider the product of
five $\mathcal{N}{=}2$ minimal models $\prod_i\mathcal{M}_{p_i}$ with central charge
\begin{equation}\label{eq:cmm}
  c = \sum_{i=1}^5 c_i = \sum_{i=1}^5\frac{3p_i}{p_i+2},\qquad p_i\in\mathbb{Z}_{>0}.
\end{equation}

Such a product of minimal models is known to correspond to a Berglund--H\"{u}bsch
Calabi--Yau manifold (or its orbifold) defined by the Fermat polynomial
$W_0(y)=\sum_i y_i^{p_i+2}$.

Primary fields of $\mathcal{N}{=}2$ minimal models are labelled by three integers $(l,q,s)$ with
\begin{equation}
  0\leq l\leq p;\quad -l\leq q\leq l;\quad s\in\mathbb{Z}\mod 4;\quad
  l+q+s=0\mod 2.
\end{equation}
The primary field $\Phi^s_{l,q}(z)$ has conformal dimension and $U(1)$ charge
\begin{equation}\label{eq:DeltaQ_mm}
  \Delta = \frac{l(l+2)-q^2}{4(p+2)} + \frac{s^2}{8},\qquad
  Q = \frac{q}{p+2} - \frac{s}{2}.
\end{equation}
There is also the identification
\begin{equation}
  (l,q,s)\simeq(p-l,\,q+(p+2),\,s+2).
\end{equation}

The steps of the construction of the four-dimensional heterotic string using a product of five
$\mathcal{N}{=}2$ minimal models largely parallel the construction in Chapter~2. Similarly,
among the left massless BRST-invariant vertex operators one finds spinors corresponding to
the $\mathcal{N}{=}1$ super-Poincar\'{e} currents (their integrals satisfy the commutation
relations for supercharges). The GSO$_L$ condition then requires all left vertex operators to
be mutually local with the $\mathcal{N}{=}1$ SUSY generators. In the right sector, among the
massless BRST-invariant vertex operators one finds those corresponding to gauge-symmetry
currents, and the GSO$_R$ condition retains only right vertex operators mutually local with
them.

In this approach the complete vertex operators take the form
\begin{equation}
  V(z,\bar z) = P^L[\ldots](z)\,\Phi^{\vS^L}_{\vec{l}^L,\vec{q}^L}(z)\,
    e^{q\phi+i\vlambda\cdot\vec{\mathcal{H}}+ip_\mu x^\mu}(z)
    \otimes P^R[\ldots](\bar z)\,\Phi^{\vS^R}_{\vec{l}^R,\vec{q}^R}(\bar z)\,
    e^{i\veta\cdot\vec{Y}+i\vLambda\cdot\vec{\Phi}+ip_\mu x^\mu}(\bar z),
\end{equation}
where $\Phi^{\vS}_{\vec{l},\vec{q}}(z)=\prod_{i=1}^5(l_i,q_i,s_i)(z)$.

The selection of complete physical vertex operators as products of left and right vertex
operators requires that they enter a modular-invariant partition function. For the minimal
admissible group it takes the form \cite{Greene1990}:
\begin{equation}\label{eq:Z}
  Z(\tau,\bar\tau) = \sum_{\substack{\vec{l}^L,\vec{l}^R \\ \vec{\mu}^L-\vec{\mu}^R=
    \frac{N}{2}\vec{\beta}^{(0)}+\sum_j n_j\vec{\beta}^{(j)}+(2;\vec{0};\vec{0})}}
  A^{\vec{l}^L}_{\vec{l}^R}\,\chi^{\vec{l}^L}_{\vec{\mu}^L}\,\bar\chi^{\vec{l}^R}_{\vec{\mu}^R},
\end{equation}
where $\vec{\mu}=(\mu_0;\vec{q};\vec{s})$ (with $\mu_j=q_j$, $\mu_{j+5}=s_j$,
$j=1,\ldots,5$) and $\mu^L_0,\mu^R_0=0,1,2,3\mod 4$ correspond to
$\vlambda,\vLambda\in[0],[v],[s],[c]$. We have introduced the vector
$\vec{\beta}^{(0)}=(2,\ldots,2)$ and vectors $\vec{\beta}^{(j)}$ with all components zero
except $\beta^{(j)}_0=\beta^{(j)}_{j+5}=2$, and $N,n_j\in\mathbb{Z}$.

The simplest modular-invariant case is when $A^{\vec{l}^L}_{\vec{l}^R}=\delta_{\vec{l}^L,\vec{l}^R}$, i.e.\
$\vec{l}^L=\vec{l}^R$ for all physical states.

The partition function includes only $\vec{\mu}^L,\vec{\mu}^R$ corresponding to states
satisfying BRST, GSO$_L$ and GSO$_R$. The shift $(2;\vec{0};\vec{0})$ corresponds to the
passage from the $\mathcal{N}{=}2$ closed fermionic string theory to the heterotic string
theory via the Gepner map \cite{Gepner1989}; under this map the right sector undergoes the
replacement of $SO(1,3)$ representations by $SO(10)$ representations:
$[0]\leftrightarrow[v]$, $[s]\leftrightarrow[c]$.

The shift $\vec{\mu}^L-\vec{\mu}^R=\tfrac{N}{2}\vec{\beta}^{(0)}$ corresponds to the action
of super-Poincar\'{e} generators from the left or $E(8)\times E(6)$ generators from the
right; the shift by $\vec{\beta}^{(j)}$ allows states with $\mu^L_0-\mu^R_0+2=s^L_j-s^R_j=2$,
i.e.\ in addition to $[0]\times[v]$ and $[v]\times[0]$ representations, it allows $[0]\times[0]$
and $[v]\times[v]$ states (and analogously for $[s]$ and $[c]$).

In the orbifold case the conditions on complete physical states change according to the choice
of admissible group $G$. The sum in \eqref{eq:Z} is replaced by a sum over the following
$\vec{\mu}^L,\vec{\mu}^R$:
\begin{equation}\label{eq:orbcond}
  \sum_{j=1}^5 \frac{(\mu^L_j+\mu^R_j)\,\gamma^a_j}{p_j+2}\in 2\mathbb{Z};
  \qquad
  \vec{\mu}^L-\vec{\mu}^R = \frac{N}{2}\vec{\beta}^{(0)}
    + \sum_j n_j\vec{\beta}^{(j)}
    + \sum_a t^a(0;2\vec{\gamma}_a;\vec{0})
    + (2;\vec{0};\vec{0}),
\end{equation}
where $N,n_j,t^a\in\mathbb{Z}$ and $\vec{\gamma}_a\in G$ are the generators of $G$.
The first condition means that complete physical vertex operators are invariant under the
action of $G$ as realised by the operators \cite{Belavin2024,Greene1990}:
\begin{equation}
  \hat{g}_{\vec{\gamma}_a}\cdot V^{\vec{l}^L,\vec{l}^R}_{\vec{\mu}^L,\vec{\mu}^R}(z,\bar z)
  = \exp\!\left(i\pi\sum_{j=1}^5\frac{(\mu^L_j+\mu^R_j)\,\gamma^a_j}{p_j+2}\right)
    V^{\vec{l}^L,\vec{l}^R}_{\vec{\mu}^L,\vec{\mu}^R}(z,\bar z).
\end{equation}
The second condition in \eqref{eq:orbcond} means that the left and right parts of a complete
vertex operator may now also differ by the action of the corresponding spectral flows
$U_{\vec{\gamma}_a}$ (see \cite{Belavin2024,Greene1990} for further details).

Using the correspondence between this construction and the free-field construction, we will
be able to generalise the latter to the case of orbifolds of Berglund--H\"{u}bsch Calabi--Yau
manifolds. We verify that conditions on the complete vertex operators analogous to
\eqref{eq:orbcond} arise.

\subsection{Correspondence Between the Two Constructions}

We expect that for Berglund--H\"{u}bsch Calabi--Yau manifolds defined by a Fermat-type
polynomial there is a connection between the free-field construction of the four-dimensional
heterotic string and the minimal-model construction. In this section we establish the
correspondence between products of primary fields and Calabi--Yau vertex operators in
terms of free fields:
\begin{equation}
  \Phi^{\vS}_{\vec{l},\vec{q}}(z) = \prod_{i=1}^5(l_i,q_i,s_i)
  \;\longleftrightarrow\;
  V^{\vS}_{\vm,\vn}(z) = P_{\mathrm{CY}}[\partial X^-_i,\partial X^+_i,\partial H_i]\,
    \exp[\vm\cdot\vX^-+\vn\cdot\vX^++i\vS\cdot\vH].
\end{equation}

Comparing the central charges of the two constructions, \eqref{eq:cCY} and \eqref{eq:cmm},
gives
\begin{equation}
  p_i+2 = \frac{1}{a^+_i a^-_i} = \frac{d}{k_i}.
\end{equation}
For the Fermat case we thus obtain the well-known relation $p_i+2=A_{ii}$. Unlike the
free-field construction, which is applicable to any Berglund--H\"{u}bsch Calabi--Yau, the
minimal-model construction admits only those Berglund--H\"{u}bsch Calabi--Yau manifolds
for which $d=\sum_j k_j$ is divisible by each weight $k_i$. We consider only the Fermat
case: $d/k_i=A_{ii}$.

Comparing conformal dimensions and $U(1)$ charges in the two constructions, \eqref{eq:DeltaCY},
\eqref{eq:QCY} and \eqref{eq:DeltaQ_mm}, we obtain the relations between the labels
$(l_i,q_i,s_i)$ and the vectors $\vm,\vn,\vS$:
\begin{equation}\label{eq:correspond}
  l_i = \frac{m_i a^-_i + n_i a^+_i}{a^+_i a^-_i} = m_i + A_{ii}n_i,\qquad
  q_i = \frac{m_i a^-_i - n_i a^+_i}{a^+_i a^-_i} = m_i - A_{ii}n_i,\qquad
  s_i = -2S_i.
\end{equation}
These relations allow, for a given vertex operator $V^{\vS}_{\vm,\vn}(z)$, to identify the
product of primary fields $\Phi^{\vS}_{\vec{l},\vec{q}}(z)$ having the same conformal
dimension and $U(1)$ charge.

However, to establish the correspondence between the constructions it is also important to
identify other properties of the primary fields of $\mathcal{N}{=}2$ minimal models. For
example, unitarity gives the bound
\begin{equation}
  0 \le l_i \le p_i = A_{ii}-2.
\end{equation}
This inequality resembles the condition on admissible deformations of $W_0$ given in
\eqref{eq:defcond}. The selection from $\Delta,\Delta^*$ of vectors $\vm,\vn$ corresponding
to admissible deformations, $m_i\le A_{ii}-2$ and $A_{ii}n_i\le A_{ii}-2$, in the free-field
construction is accomplished by the requirement of Borisov cohomologies:
\begin{equation}\label{eq:Dkernel}
  D\,V^{\vS}_{\vm,\vn}(z) = 0
  \;\Rightarrow\; m_i + A_{ii}n_i \ge 0,
\end{equation}
\begin{equation}
  V^{\vS}_{\vm,\vn}(z) \ne D\,V^{(0)}(z)
  \;\Rightarrow\; m_i + A_{ii}n_i \le A_{ii}-2.
\end{equation}
These relations are obtained straightly by examining the pole structure of the integrands in the contour integrals defining the Borisov differential. Note also that vertex operators differing by a $D$-exact vertex operator are equivalent.
In the Fermat case:
\begin{equation}
  \begin{cases}
    \vm_1 = \vm_2 + \vec{u}_i\\
    \vn_1 = \vn_2 - \vec{v}_i\\
    S_{1j}-S_{2j} = -2\delta_{ij}
  \end{cases}
  \;\Rightarrow\; V_1(z) - V_2(z) = D\cdot V^{(0)}(z),
\end{equation}
where $(\vec{u}_i)_j=A_{ij}=A_{ii}\delta_{ij}$, $(\vec{v}_i)_j=\delta_{ij}$. The equivalence of
such vertex operators corresponds to the symmetry $q_i\sim q_i+2(p_i+2)$, with $p_i+2=A_{ii}$.

The conditions on complete vertex operators following from modular invariance of the
partition function \eqref{eq:Z} in the case $A^{\vec{l}^L}_{\vec{l}^R}=\delta_{\vec{l}^L,\vec{l}^R}$,
using the relations \eqref{eq:correspond}, take the form:
\begin{equation}
  \vm^L = \vm^R + \frac{N}{2}\vec{a}^+,\qquad
  \vn^L = \vn^R - \frac{N}{2}\vec{a}^-,\quad N\in\mathbb{Z}.
\end{equation}

These are exactly the constraints we arrive at in the free-field construction in Section~2. This fact verifies the modular invariance of the free-field construction in the Fermat case for the minimal admissible group. 

Analogously, for orbifolds, modular invariance together with the relations \eqref{eq:orbcond} will imply the following:
\begin{equation}\label{eq:mnLRorb}
  \vm^L = \vm^R + \frac{N}{2}\veca^+ - \vw,\qquad
  \vn^L = \vn^R - \frac{N}{2}\veca^- + \hat{B}\vw,\quad N\in\mathbb{Z},\;\vw\in G.
\end{equation}

This implies an important consequence: the vectors $\vm^L$ and $\vm^R$, as well as the vectors $\vn^L$ and $\vn^R$, will belong to different lattices. As we will establish by defining the lattices $M(G)$ and $N(G)$ for an arbitrary admissible group $G$, the relation \eqref{eq:mnLRorb} provides the key intuition: if the right vertex algebra of the heterotic string is determined by the chosen admissible group $G$ ($\vm^R\in M(G)$, $\vn^R\in N(G)$), then the left vertex algebra must be determined by the minimal admissible group ($\vm^L\in M(G^{min}_{adm})$, $\vn^L\in N(G^{min}_{adm})$).

In the next section, we discuss the generalisation of the free-field construction to orbifolds of Berglund--H\"{u}bsch type Calabi--Yau. Unlike the case of the minimal admissible group, we find that the resulting theory is not merely a hybrid of two conformal field theories, but also a hybrid of two Calabi--Yau manifolds: the original Calabi--Yau manifold as a hypersurface in the left sector and its orbifold by the chosen admissible group $G$ in the right sector. Complete vertex operators are constructed as products of left and right vertices subject to an additional condition, which turns out to be equivalent to the simplest case of modular invariance in the minimal-model construction. It is precisely this construction that yields the correct number of massless singlet vertex operators as well as $27$- and $\overline{27}$-representations of $E(6)$, ensuring an exact correspondence between the obtained vertex operators and the same vertices in the minimal-model construction.
\section{Free-Field Construction for Calabi--Yau Orbifolds}
\label{sec:orbifolds}

In this section, guided by the correspondence between the minimal-model and free-field
constructions, we discuss the generalisation of the latter to Berglund--H\"{u}bsch Calabi--Yau
orbifolds. It is important to note that a key advantage of the free-field construction over the
minimal-model construction is that it applies to all Berglund--H\"{u}bsch Calabi--Yau manifolds. 

\subsection{Combinatorial Data for Orbifolds}
From the orbifold data given by the matrix $A_{ij}$ and the admissible group $G$, we can, as
in Section~2, obtain the Batyrev--Borisov combinatorial data by constructing the dual lattices
$M(G)$ and $N(G)$. This is done similarly to \cite{Borisov2004, Borisov2011}. The dual lattices are now defined as follows: the lattice $N(G)$ is an extension of
$N_0$ by the vector $\vec{a}^-$ and by vectors corresponding to the generators of $G$:
\begin{equation}\label{eq:NG}
  N(G) = N_0 + \mathbb{Z}\,\vec{a}^- + \sum_a\mathbb{Z}\,\hat{B}{\vec{\gamma}}_a,
\end{equation}
where $B_{ij}=(A^{-1})_{ij}$, so $(\hat{B}{\vec{\gamma}}_a)_i=\sum_j B_{ij}\gamma^a_j$. Here, $\vec{\gamma}^a$ corresponds to a following generator of an admissible group $G$:
\begin{equation}
    y_i\;\longrightarrow\; y_i\exp\!\left(\sum_{j=1}^5 B_{ij}\gamma^a_j\right),\,\,\,\gamma^a_i\in\mathbb{Z},
\end{equation}
which implies that $\vec{a}^-$ corresponds to the generator of the minimal admissible group.

We then find five vectors $\vec{e}^*_\beta$ forming a basis of $N(G)$, and the basis
$\vec{e}_\alpha$ of the dual lattice $M(G)$ is determined by
\begin{equation}
  \vec{e}_\alpha\cdot\vec{e}^*_\beta = \delta_{\alpha\beta}.
\end{equation}

Duality implies that $M(G)$ is an extension of $M_0$ by the generators $\vec{\gamma}^{*b}$
of the mirror group $G^*$:
\begin{equation}\label{eq:MG}
  M(G) = M_0 + \mathbb{Z}\,\vec{a}^+ + \sum_b\mathbb{Z}\,\vec{\gamma}^{*b}.
\end{equation}

The Batyrev dual polytopes $\Delta$ and $\Delta^*$ are defined as in Section~2:
\begin{align}
  \Delta   &= \{\vm\in M(G)\mid\vm\in K,\;\vm\cdot\vec{a}^-=1\},\\
  \Delta^* &= \{\vn\in N(G)\mid\vn\in K^*,\;\vec{a}^+\cdot\vn=1\}.
\end{align}

Points $\vm\in\Delta$ satisfying $m_i\le A_{ii}-2$ (Fermat case) are admissible deformation
vectors, while points $\vn\in\Delta^*$ satisfying $A_{ii}n_i\le A_{ii}-2$ are related to
mirror deformations via $n^{\mathrm{def}}_i=A_{ii}n_i$.

For some classes of orbifolds the numbers of deformations and mirror deformations coincide
with the Hodge numbers $h^{2,1}$ and $h^{1,1}$, respectively, and the vertex operators of
the 27 and $\overline{27}$ supermultiplets are then parametrised by deformation vectors.
However, for certain orbifolds there are more 27 or $\overline{27}$ vertex operators than
corresponding deformations. In that case the numbers of generations and anti-generations
obtained in the construction described below do coincide with the Hodge numbers, which for
such orbifolds are not equal to the number of deformations.  In fact, the full Hodge numbers in such specific cases are computed as the number of admissible (or mirror) deformations plus the number of special pairs of elements belonging to the mirror groups $G$ and $G^*$. This issue, along with the refined formula for the Hodge numbers of such exceptional Calabi--Yau orbifolds, will be discussed in detail in an upcoming paper by S.~Aleshin, A.~Belavin, and G.~Koshevoy.

\subsection{Left and right vertex algebras}
Having defined the combinatorial data for the orbifold case, we will now use it to construct the free-field realisation of the four-dimensional heterotic string compactified on a Berglund--H\"{u}bsch type Calabi--Yau orbifold. We propose the following construction: we first define the left and right vertex algebras separately, and then construct the complete vertex operators as tensor products of left and right states, subject to an additional pairing condition.

Starting with the right sector, the vertex algebra $\mathrm{VA}^R(G)$ consists of operators of the form:
\begin{equation}\label{eq:VRorb}
  V^R_{\veta,\vLambda,\vm^R,\vn^R,\vS^R}(\bar z)
  = P(\bar\partial^k\bar x^a,\bar\partial^l Y^I,\bar\partial^{l'}\Phi^\alpha,
       \bar\partial^r H_i,\bar\partial^s\bar X^+_i,\bar\partial^t\bar X^-_i)\,
    e^{i\veta\cdot\vec{Y}+i\vLambda\cdot\vec{\Phi}+\vm^R\cdot\vX^-+\vn^R\cdot\vX^++i\vS^R\cdot\vH+ip_\mu x^\mu},
\end{equation}
which must satisfy the following requirements. First, they must represent non-trivial BRST and Borisov cohomology classes. Second, the vectors $\vm^R$ and $\vn^R$ must belong to the lattices determined by the admissible group $G$, defined in \eqref{eq:NG} and \eqref{eq:MG}: $\vm^R\in M(G)$ and $\vn^R\in N(G)$ in the NS sector; or $\vm^R\in M(G)\pm\tfrac{1}{2}\veca^+$ and $\vn^R\in N(G)\pm\tfrac{1}{2}\veca^-$ if the Calabi--Yau sector is in the Ramond (R) representation of the super-Virasoro algebra.

 Among the massless right vertex operators, following \cite{Belavin2025}, we find the currents of the $E(8)$ and $SO(10)$ algebras \eqref{eq:E8}, \eqref{eq:SO10}, as well as the $U(1)$ current \eqref{eq:U1} originating from the Calabi--Yau subsector. In addition, we find vertex operators \eqref{eq:SO10spinors} corresponding to the $\mathbf{16}$ and $\overline{\mathbf{16}}$ spinor representations of $SO(10)$. The inclusion of these representations together with the $SO(10)$ and $U(1)$ currents extends the gauge symmetry algebra to $E(6)$. We then require all right vertex operators to be mutually local with these gauge currents, i.e., we impose the $\mathrm{GSO}^R$ condition. The duality of the lattices $M(G)$ and $N(G)$, combined with the $\mathrm{GSO}^R$ condition, ensures that all such right vertex operators are mutually local with each other.

It is precisely the right vertex algebra $\mathrm{VA}^R(G)$ that encodes all the information associated with the chosen admissible group $G$, such as the number of singlets, generations, and anti-generations. For example, consider the quintic orbifold ($W_0=\sum_i y_i^5$) with the admissible group generated by the following elements:
\begin{equation}
    G = <[1,1,1,1,1],\, [0,1,1,4,4] >.
\end{equation}
Here, we find a total of $468=234+234$ right singlet vertex operators. These ultimately combine to form $234$ complete singlet vertex operators and $234$ vertex operators corresponding to their antiparticles (see the Appendix for details). 

Next, we define the left vertex algebra $\mathrm{VA}^L(G^{\min}_{\mathrm{adm}})$. A general left vertex operator takes the form:
\begin{equation}\label{eq:VLorb}
  V^L_{q,\vlambda,\vm^L,\vn^L,\vS^L}(z)
  = P(\partial^k x^\mu,\partial^l \mathcal{H}_a,\partial^r H_i,\partial^s X^+_i,\partial^t X^-_i)\,
    e^{q\phi(z)}\,
    e^{i\vlambda\cdot\vec{\mathcal{H}}(z)+\vm^L\cdot\vX^-+\vn^L\cdot\vX^++i\vS^L\cdot\vH+ip_\mu x^\mu},
\end{equation}
and these operators must satisfy several conditions. First, they must also correspond to non-trivial BRST and Borisov cohomology classes  and be  $\mathcal{N}=1$ super-Poincar\'{e} covariant.

Since the right-moving parts of the physical operators are mutually local, their left-moving counterparts must also be mutually local among themselves to ensure the mutual locality of the complete vertex operators. Consequently, the vectors $\vm^L$ and $\vn^L$ must belong to the dual lattices. We restrict them to the lattices constructed from the minimal admissible group, exactly as in \eqref{eq:NGmin}: $\vm^L\in M(G^{\min}_{\mathrm{adm}})$ and $\vn^L\in N(G^{\min}_{\mathrm{adm}})$ in the NS sector; or $\vm^L\in M(G^{\min}_{\mathrm{adm}})\pm\tfrac{1}{2}\veca^+$ and $\vn^L\in N(G^{\min}_{\mathrm{adm}})\pm\tfrac{1}{2}\veca^-$ in the R sector.

This unconventional choice is motivated by two reasons. First, the correspondence with the Gepner construction yields the relation between the left and right vectors given in \eqref{eq:mnLRorb}. It is straightforward to see from the definitions \eqref{eq:NG} and \eqref{eq:MG} that this relation forces $\vm^L \in M(G^{\min}_{\mathrm{adm}})$ and thus, by duality of lattices, $\vn^L \in N(G^{\min}_{\mathrm{adm}})$ --- this means that such choise is imposed by the modular invariance. Second, by applying the correspondence \eqref{eq:correspond} to explicitly known states, such as the singlets and the $27$- and $\overline{27}$-representations, we observe that their left-moving (holomorphic) parts, when translated into the free-field language, naturally fall into the $\vm^L\in M(G^{\min}_{\mathrm{adm}})$, $\vn^L\in N(G^{\min}_{\mathrm{adm}})$ classification (as demonstrated in the Appendix).

Finally, the left vertex operators \eqref{eq:VLorb} must satisfy the $\mathrm{GSO}^L$ condition, meaning they must be mutually local with the $\mathcal{N}{=}1$ super-Poincar\'{e} currents \eqref{eq:J}.

\subsection{Complete Vertex Algebra for the Orbifold Case}
The most important step is the construction of complete vertex operators as products of left and right parts. We have already seen this for the case of the minimal admissible group in Section~\ref{sec:Gmin}, where we considered only vertex operators that are diagonal with respect to the Calabi--Yau part ($\vm^L=\vm^R$, $\vn^L=\vn^R$), along with their superpartners and gauge-group partners. Ultimately, this led to the following constraints:
\begin{equation}\label{eq:GminMNquasidiag}
  \vm^L = \vm^R + \frac{N}{2}\vec{a}^+,\qquad
  \vn^L = \vn^R - \frac{N}{2}\vec{a}^-,\quad N\in\mathbb{Z}.
\end{equation}

Now, for the orbifold case, we propose the following pairing conditions: we consider complete vertex operators of the form
\begin{equation}
  V(z,\bar z) = V^L_{q,\vlambda,\vm^L,\vn^L,\vS^L}(z)
    \otimes V^R_{\veta,\vLambda,\vm^R,\vn^R,\vS^R}(\bar z),
\end{equation}
such that the left and right parts individually satisfy the  above-mentioned  conditions. We allow only those complete vertex operators for which the following holds:
\begin{equation}\label{eq:m+An}
    \vm^L+\hat{A}\vn^L = \vm^R+\hat{A}\vn^R,
\end{equation}
where $\hat{A}$ is the degree matrix, such that $(\hat{A}\vn)_i=\sum_j A_{ij}n_j$, and in the Fermat case we have:
\begin{equation}
    m^L_i +A_{ii}n_i^L = m_i^R +A_{ii}n_i^R.
\end{equation}

For the minimal admissible group, it is easy to verify that the relation \eqref{eq:GminMNquasidiag} also follows from the condition \eqref{eq:m+An}. 

By restricting ourselves specifically to the set of complete vertex operators satisfying \eqref{eq:m+An}, in the Fermat case we obtain the correct number of massless complete singlet vertex operators, as well as $27$- and $\overline{27}$-supermultiplets. Moreover, in the Appendix, using an orbifold of the quintic as an example, we illustrate that by means of \eqref{eq:correspond} there is a one-to-one correspondence between the states obtained in this way and the states obtained in the Gepner construction. This confirms the modular invariance of the thus-constructed four-dimensional heterotic string theory for the Fermat case.

Although the condition \eqref{eq:m+An} was derived experimentally from the requirement of matching the massless spectrum, it has a direct connection to modular invariance. Indeed, the partition function of the Gepner construction \eqref{eq:Z} (or its variation for the orbifold case) includes the coefficients $A^{\vec{l}^L}_{\vec{l}^R}$. As mentioned in Section~\ref{subsec:minmod}, the simplest case satisfying modular invariance corresponds to $A^{\vec{l}^L}_{\vec{l}^R} = \delta_{\vec{l}^L, \vec{l}^R}$, which imposes the condition $\vec{l}^L=\vec{l}^R$ on the complete vertex operators in the Gepner construction.

Using the correspondence between the constructions \eqref{eq:correspond}, we find that this leads to the condition \eqref{eq:m+An}, which once again confirms the modular invariance of the proposed construction for a Calabi--Yau manifold defined by a Fermat-type polynomial.

\subsection{Singlets and $27$-, $\overline{27}$-Representations in the Orbifold Case}

Thus, when constructing complete physical vertex operators that satisfy the listed conditions, we are primarily interested in the massless complete vertex operators of the $27$- and $\overline{27}$-representations of $E(6)$, as well as massless singlets. In this section, we provide the definitions of these vertex operators and highlight certain specific features found in the case of Calabi--Yau orbifolds.

The vertex operators of the $27$-supermultiplet take the following form:
\begin{equation}\label{eq:general27}
  e^{-\phi(z)}\cdot \mathcal{V}^{(Q=+1,\Delta=1/2)}_{CY}(z) \otimes e^{i\vLambda\cdot\vec{\Phi}(\bar{z})}\cdot\mathcal{V}^{(Q=+1,\Delta=1/2)}_{CY}(\bar{z}),
\end{equation}
and the $\overline{27}$ vertex operators:
\begin{equation}\label{eq:general27bar}
  e^{-\phi(z)}\cdot \mathcal{V}^{(Q=+1,\Delta=1/2)}_{CY}(z) \otimes e^{i\vLambda\cdot\vec{\Phi}(\bar{z})}\cdot \mathcal{V}^{(Q=-1,\Delta=1/2)}_{CY}(\bar{z}),
\end{equation}
plus their $E(6)$ partners and superpartners. Here, $\mathcal{V}^{(Q,\Delta)}_{CY}(z)$ denotes a Calabi--Yau vertex operator with the corresponding $U(1)$ charge and conformal dimension. Recall that the complete vertex operators must satisfy: BRST, Borisov cohomologies, $\mathrm{GSO}^{L,R}$, as well as the condition on the product of the left and right vertex operators \eqref{eq:m+An}.

In standard orbifold cases, these vertex operators take the form:
\begin{equation}\label{eq:simple27}
  e^{-\phi(z)}\exp[\vm\cdot\vX^-](z)\otimes\exp[\vm\cdot\vX^-+i\vLambda\cdot\vec{\Phi}](\bar z)
\end{equation}
for the $27$-representation, where the vectors $\vm\in M(G)\subseteq M(G^{\min}_{\mathrm{adm}})$ are the admissible deformation vectors, and
\begin{equation}\label{eq:simple27bar}
  e^{-\phi(z)}\exp[\vec{m}\cdot\vX^-](z)\otimes\exp[\vec{n}\cdot\vX^++i\vLambda\cdot\vec{\Phi}](\bar z),\,\,\vm = \hat{A}\vn
\end{equation}
for the $\overline{27}$-representation, where $\vm\in M(G^{\min}_{\mathrm{adm}})$, $\vn\in N(G)$, and the vectors $\vm=\hat{A}\vn$ are mirror deformation vectors. In this case, the correct number of types of chiral massless matter is ensured by the equality between the Hodge numbers $h^{2,1}$, $h^{1,1}$ of the chosen orbifold and the number of its deformations and mirror deformations, respectively.

However, there are also special cases in which the Hodge numbers do not coincide with the number of admissible deformations. In such cases, we find additional generation vertex operators of the general form \eqref{eq:general27} and/or \eqref{eq:general27bar}, whose right parts are parametrised by a pair of vectors $\vm$, $\vn$ associated with special pairs of elements of the groups $G$ and $G^*$.

Finally, we define the massless singlet vertex operators as follows:
\begin{equation}\label{eq:generalsinglet}
  e^{-\phi(z)}\cdot \mathcal{V}^{(Q=+1,\Delta=1/2)}_{CY}(z) \otimes \mathcal{V}^{(Q=0,\Delta=1)}_{CY}(\bar{z}).
\end{equation}

 One of the key results of this work is the explicit derivation of the complete list of massless singlet vertex operators for the quintic orbifold with Hodge numbers $(17,21)$ (see Appendix). To verify the validity of the free-field construction in the orbifold case, we independently computed the spectrum of massless singlets for this example in the minimal-model framework. We find perfect agreement between the results obtained via the two constructions.

Within the free-field construction, we identify the singlet states using the following algorithm.

First, we determine all possible massless right-moving singlet vertex operators that are neutral under the $E(8)\times E(6)$ gauge group. These operators consist entirely of the Calabi--Yau part, as shown in \eqref{eq:generalsinglet}, and must satisfy the following conditions: BRST, Borisov cohomologies, $\mathrm{GSO}^R$. They must have conformal dimension $\Delta^R_{\mathrm{CY}}=1$ and $U(1)$ charge $Q^R_{\mathrm{CY}}=0$. Furthermore, the vectors $\vm^R$ and $\vn^R$ appearing in these right-moving vertices must belong to the lattices determined by the admissible group $G$, namely $\vm^R\in M(G)$ and $\vn^R\in N(G)$. For the $(17,21)$ example, we find a total of $468=234+234$ right-moving massless singlets. 

Next we find massless left-handed vertices, which are spacetime scalars (i.e., singlets with respect to $SO(1,3)$). These must possess conformal dimension $\Delta^L_{\mathrm{CY}}=1/2$ and $U(1)$ charge $Q^L_{\mathrm{CY}}=+1$. They must satisfy BRST and $\mathrm{GSO}^L$, and belong to Borisov cohomologies, and their defining vectors must satisfy $\vm^L\in M(G^{\min}_{\mathrm{adm}})$ and $\vn^L\in N(G^{\min}_{\mathrm{adm}})$.

From these left- and right-moving components, we construct complete vertex operators which satisfy the left-right pairing condition \eqref{eq:m+An}. In the $(17,21)$ case, only half of the identified right-moving massless singlets can be successfully paired to form complete singlet vertices. Consequently, we obtain exactly $234$ singlets for this model (see the Appendix for details). All of these vertices satisfy the following:
\begin{equation}
    m_i^L = m_i^R +A_{ii}n_i^R.
\end{equation}

The remaining $234$ right-moving singlets pair by the relations \eqref{eq:m+An} with left-moving vertices of an opposite chirality (having a left $U(1)$ charge $Q^L_{\mathrm{CY}}=-1$), which correspond to the antiparticles of the original set.

It is important to emphasize that while the left-moving components of these singlet vertices can repeat, their right-moving counterparts are strictly unique. It is precisely the right vertex algebra $\mathrm{VA}^R(G)$ that encodes all the information regarding the number of singlets (as well as the number of generations) dictated by the choice of the admissible group $G$.

The algorithm for constructing the ${27}$ and $\overline{{27}}$ vertex operators follows a similar logic. The only difference lies in the right-moving part, which must instead belong to the vector representation of $SO(10)$ and have a conformal dimension $\Delta^R_{\mathrm{CY}}=1/2$. The $U(1)$ charge is $Q^R_{\mathrm{CY}}=+1$ for the ${27}$ generations, and $Q^R_{\mathrm{CY}}=-1$ for the $\overline{{27}}$ anti-generations.
\section{Conclusion}

In the present paper we have discovered a connection between two constructions of the
Calabi--Yau sector of the four-dimensional heterotic string, by establishing a correspondence
between primary fields of $\mathcal{N}{=}2$ SCFT minimal models and vertex operators in
terms of free fields. This connection first confirms modular invariance of the free-field
construction for the case where the admissible group in the definition of the
Berglund--H\"{u}bsch Calabi--Yau is chosen to be minimal.

Second, using this correspondence we were able to generalise the free-field construction to
the case of Berglund--H\"{u}bsch Calabi--Yau orbifolds. In the Fermat case we find
conditions on the complete physical vertex operators analogous to those in the minimal-model
construction.

Complete physical vertex operators are constructed from products of left and right parts such
that each part individually satisfies the BRST, Borisov cohomology, and GSO conditions. It turns out that in the case of compactification on a Calabi--Yau orbifold, the left vertex operators are constructed based on the combinatorial data obtained for the minimal admissible group, whereas the right ones are based on the dual lattices constructed from the given admissible group $G$. 

Nevertheless, this hybrid nature of the theory leads to the correct result, consistent with the Hodge numbers of the given orbifold and with the massless spectrum obtained from the minimal-model construction. Moreover, the additional conditions imposed on the set of complete vertex operators---formed as products of the left and right parts---correspond to one of the possible solutions to the modular invariance requirements known from the Gepner construction.

The free-field construction, unlike the minimal-model construction, is applicable to all
Berglund--H\"{u}bsch Calabi--Yau manifolds. It would be interesting to consider an even
wider class---toric Calabi--Yau manifolds, whose definition was proposed in the works of
Batyrev. We expect that a generalisation of the free-field construction to this class of
Calabi--Yau manifolds is possible.

\section*{Acknowledgements}
Authors acknowledge A.~Litvinov, S.~Aleshin and G.~Koshevoy for helpful and
interesting discussions. The work is supported by  the Russsian Science Foundation  grant 23-12-00333.
\newpage
\section*{Appendix: Singlets and $27$-/$\overline{27}$-Vertex Operators\\
for the Quintic Orbifold $(17,21)$}
\addcontentsline{toc}{section}{Appendix: Singlets and 27-/\protect$\overline{27}$-Vertex Operators for the Quintic Orbifold (17,21)}

As an example, let us consider the quintic orbifold ($A_{ii}=5$) with an admissible group generated by:
\begin{equation}
    G=<[1,1,1,1,1],[0,1,1,4,4]>.
\end{equation}

Its mirror group then has the following generators:
\begin{equation}
    G^*=<[1,1,1,1,1],[0,1,4,0,0],[0,0,0,1,4]>.
\end{equation}

According to \eqref{eq:NG}, the lattice $N(G)$ is defined as
\begin{equation}
    N(G)=N_0+\mathbb{Z}\vec{a}^-+\mathbb{Z}\left(0,\dfrac{1}{5},\dfrac{1}{5},\dfrac{4}{5},\dfrac{4}{5}\right).
\end{equation}

We obtain the following basis of the lattice $N(G)$:
\begin{equation}
    \begin{array}{c}
        \vec{e}^*_1=\left(\dfrac{1}{5},\dfrac{2}{5},\dfrac{2}{5},0,0\right),
        \quad\vec{e}^*_2=(0,1,0,0,0),\\
        \vec{e}^*_3=(0,0,1,0,0),
        \quad\vec{e}^*_4=(0,0,0,1,0),\\
        \vec{e}^*_5=\left(\dfrac{1}{5},\dfrac{1}{5},\dfrac{1}{5},\dfrac{1}{5},\dfrac{1}{5}\right).
    \end{array}
\end{equation}

The basis $\vec e_\alpha$ of the dual lattice $M(G)$ is determined by the condition $\vec e_\alpha \vec e_\beta^*=\delta_{\alpha\beta}$:
\begin{equation}
    \begin{array}{c}
        \vec{e}_1=(5,0,0,0,-5),
        \quad\vec{e}_2=(-2,1,0,0,1),\\
        \vec{e}_3=(-2,0,1,0,1),
        \quad\vec{e}_4=(0,0,0,1,-1),\\
        \vec{e}_5=(0,0,0,0,5).
    \end{array}
\end{equation}

We also define the dual lattices corresponding to the minimal admissible group as in \eqref{eq:NGmin}:
\begin{equation}
    N(G^{min}_{adm})=N_0+\mathbb{Z}\vec{a}^-,
\end{equation}
so that the basis vectors of $N(G^{min}_{adm})$ are following:
\begin{equation}
    \begin{array}{c}
        \vec{f}^*_1=(1,0,0,0,0),
        \quad\vec{f}^*_2=(0,1,0,0,0),\\
        \vec{f}^*_3=(0,0,1,0,0),
        \quad\vec{f}^*_4=(0,0,0,1,0),\\
        \vec{f}^*_5=\left(\dfrac{1}{5},\dfrac{1}{5},\dfrac{1}{5},\dfrac{1}{5},\dfrac{1}{5}\right),
    \end{array}
\end{equation}
and then the dual lattice $M(G^{min}_{adm})$ has the following bases:
\begin{equation}
    \begin{array}{c}
        \vec{f}_1=(1,0,0,0,-1),
        \quad\vec{f}_2=(0,1,0,0,1),\\
        \vec{f}_3=(0,0,1,0,1),
        \quad\vec{f}_4=(0,0,0,1,-1),\\
        \vec{f}_5=(0,0,0,0,5).
    \end{array}
\end{equation}

The 17 vertices of generations then take the following form:
\begin{equation}\label{27(17,21)}
    \exp(-\phi+\vec{m}\vec{X}^-)(z)\otimes \exp(\vec{m}\vec{X}^-+i\vec{\Lambda}\vec{\Phi}), \,\,\vec{\Lambda}\in [v],
\end{equation}
where the vectors $\vec{m}$ correspond to deformations of $W_0$, invariant under $G$:
\begin{equation}
    \begin{array}{l}
        \vec{m}= (3,1,0,0,1),(3,0,1,0,1),(3,1,0,1,0),(3,0,1,1,0),\\
        (1,2,0,0,2),(1,0,2,0,2),(1,0,2,2,0),(1,2,0,2,0),\\
        (1,0,2,1,1),(1,1,1,0,2),(1,1,1,2,0),(1,2,0,1,1),(1,1,1,1,1),\\
        (0,0,0,2,3),(0,0,0,3,2),(0,3,2,0,0),(0,2,3,0,0).
    \end{array}
\end{equation}

The remaining vertices of the 27-supermultiplet are obtained by acting on (\ref{27(17,21)}) with the $E(6)$ currents.

The 21 anti-generations vertices include five vertices corresponding to mirror deformations. Below, we list them with the corresponding products of primary fields $(l_i,q_i,s_i)$:

\begin{table}[H]
\centering
{\small
\renewcommand{\arraystretch}{1.6} 
\setlength{\tabcolsep}{8pt}

\begin{tabular}{|c|c|}
\hline
\textbf{Left-moving part ($V^L$)} & \textbf{Right-moving part ($V^R$)} \\ \hline\hline

$\begin{array}{c}
\exp(-\phi+\vec m \cdot \vec X^-) \\
\vec m=(1,0,0,2,2) \\
(1,1,0)(0,0,0)(0,0,0)(2,2,0)(2,2,0)
\end{array}$
&
$\begin{array}{c}
\exp(\vec n \cdot \vec X^+ + i\vec\Lambda \cdot \vec\Phi) \\
\vec n=\left(\dfrac15,0,0,\dfrac25,\dfrac25\right) \\
(1,-1,0)(0,0,0)(0,0,0)(2,-2,0)(2,-2,0)
\end{array}$
\\ \hline

$\begin{array}{c}
\exp(-\phi+\vec m \cdot \vec X^-) \\
\vec m=(1,1,1,1,1) \\
(1,1,0)(1,1,0)(1,1,0)(1,1,0)(1,1,0)
\end{array}$
&
$\begin{array}{c}
\exp(\vec n \cdot \vec X^+ + i\vec\Lambda \cdot \vec\Phi) \\
\vec n=\left(\dfrac15,\dfrac15,\dfrac15,\dfrac15,\dfrac15\right) \\
(1,-1,0)(1,-1,0)(1,-1,0)(1,-1,0)(1,-1,0)
\end{array}$
\\ \hline

$\begin{array}{c}
\exp(-\phi+\vec m \cdot \vec X^-) \\
\vec m=(1,2,2,0,0) \\
(1,1,0)(2,2,0)(2,2,0)(0,0,0)(0,0,0)
\end{array}$
&
$\begin{array}{c}
\exp(\vec n \cdot \vec X^+ + i\vec\Lambda \cdot \vec\Phi) \\
\vec n=\left(\dfrac15,\dfrac25,\dfrac25,0,0\right) \\
(1,-1,0)(2,-2,0)(2,-2,0)(0,0,0)(0,0,0)
\end{array}$
\\ \hline

$\begin{array}{c}
\exp(-\phi+\vec m \cdot \vec X^-) \\
\vec m=(3,0,0,1,1) \\
(3,3,0)(0,0,0)(0,0,0)(1,1,0)(1,1,0)
\end{array}$
&
$\begin{array}{c}
\exp(\vec n \cdot \vec X^+ + i\vec\Lambda \cdot \vec\Phi) \\
\vec n=\left(\dfrac35,0,0,\dfrac15,\dfrac15\right) \\
(3,-3,0)(0,0,0)(0,0,0)(1,-1,0)(1,-1,0)
\end{array}$
\\ \hline

$\begin{array}{c}
\exp(-\phi+\vec m \cdot \vec X^-) \\
\vec m=(3,1,1,0,0) \\
(3,3,0)(1,1,0)(1,1,0)(0,0,0)(0,0,0)
\end{array}$
&
$\begin{array}{c}
\exp(\vec n \cdot \vec X^+ + i\vec\Lambda \cdot \vec\Phi) \\
\vec n=\left(\dfrac35,\dfrac15,\dfrac15,0,0\right) \\
(3,-3,0)(1,-1,0)(1,-1,0)(0,0,0)(0,0,0)
\end{array}$
\\ \hline

\end{tabular}
}
\end{table}

The remaining 16 vertices of anti-generations have the following form:

\begin{table}[H]
\centering
{\small
\renewcommand{\arraystretch}{1.6}
\setlength{\tabcolsep}{8pt}

\begin{tabular}{|c|c|}
\hline
\textbf{Left-moving part ($V^L$)} & \textbf{Right-moving part ($V^R$)} \\ \hline\hline

$\begin{array}{c}
\exp(-\phi+\vec{m}\cdot\vec{X}^-) \\ 
\vec{m}=(2,0,3,0,0) \\
(2,2,0)(0,0,0)(3,3,0)(0,0,0)(0,0,0)
\end{array}$
&
$\begin{array}{c}
\exp(\vec{m}\cdot\vec{X}^-+\vec{n}\cdot\vec{X}^++i\vec{S}\cdot\vec{H}+i\vec{\Lambda}\cdot\vec{\Phi}) \\
\vec{m}=(0,1,-1,0,0),\ \vec{n}=\left(\dfrac{2}{5},-\dfrac{1}{5},\dfrac{4}{5},0,0\right),\ \vec{S}=(0,-1,1,0,0) \\
(2,-2,0)(0,2,2)(3,-5,2)(0,0,0)(0,0,0)
\end{array}$
\\ \hline

$\begin{array}{c}
\exp(-\phi+\vec{m}\cdot\vec{X}^-) \\ 
\vec{m}=(0,0,3,1,1) \\
(0,0,0)(0,0,0)(3,3,0)(1,1,0)(1,1,0)
\end{array}$
&
$\begin{array}{c}
\exp(\vec{m}\cdot\vec{X}^-+\vec{n}\cdot\vec{X}^++i\vec{S}\cdot\vec{H}+i\vec{\Lambda}\cdot\vec{\Phi}) \\
\vec{m}=(0,1,-1,0,0),\ \vec{n}=\left(0,-\dfrac{1}{5},\dfrac{4}{5},\dfrac{1}{5},\dfrac{1}{5}\right),\ \vec{S}=(0,-1,1,0,0) \\
(0,0,0)(0,2,2)(3,-5,2)(1,-1,0)(1,-1,0)
\end{array}$
\\ \hline

$\begin{array}{c}
\exp(-\phi+\vec{m}\cdot\vec{X}^-) \\ 
\vec{m}=(2,1,2,0,0) \\
(2,2,0)(1,1,0)(2,2,0)(0,0,0)(0,0,0)
\end{array}$
&
$\begin{array}{c}
(\partial X^+_3 + i\partial H_3)
\exp(\vec{m}\cdot\vec{X}^-+\vec{n}\cdot\vec{X}^++i\vec{S}\cdot\vec{H}+i\vec{\Lambda}\cdot\vec{\Phi}) \\
\vec{m}=(0,2,-2,0,0),\ \vec{n}=\left(\dfrac{2}{5},-\dfrac{1}{5},\dfrac{4}{5},0,0\right),\ \vec{S}=(0,-1,1,0,0) \\
(2,-2,0)(1,3,2)(2,-6,2)(0,0,0)(0,0,0)
\end{array}$
\\ \hline

$\begin{array}{c}
\exp(-\phi+\vec{m}\cdot\vec{X}^-) \\ 
\vec{m}=(0,1,2,1,1) \\
(0,0,0)(1,1,0)(2,2,0)(1,1,0)(1,1,0)
\end{array}$
&
$\begin{array}{c}
(\partial X^+_3 + i\partial H_3)
\exp(\vec{m}\cdot\vec{X}^-+\vec{n}\cdot\vec{X}^++i\vec{S}\cdot\vec{H}+i\vec{\Lambda}\cdot\vec{\Phi}) \\
\vec{m}=(0,2,-2,0,0),\ \vec{n}=\left(0,-\dfrac{1}{5},\dfrac{4}{5},\dfrac{1}{5},\dfrac{1}{5}\right),\ \vec{S}=(0,-1,1,0,0) \\
(0,0,0)(1,3,2)(2,-6,2)(1,-1,0)(1,-1,0)
\end{array}$
\\ \hline

$\begin{array}{c}
\exp(-\phi+\vec{m}\cdot\vec{X}^-) \\ 
\vec{m}=(2,2,1,0,0) \\
(2,2,0)(2,2,0)(1,1,0)(0,0,0)(0,0,0)
\end{array}$
&
$\begin{array}{c}
(\partial X^+_2 + i\partial H_2)
\exp(\vec{m}\cdot\vec{X}^-+\vec{n}\cdot\vec{X}^++i\vec{S}\cdot\vec{H}+i\vec{\Lambda}\cdot\vec{\Phi}) \\
\vec{m}=(0,-2,2,0,0),\ \vec{n}=\left(\dfrac{2}{5},\dfrac{4}{5},-\dfrac{1}{5},0,0\right),\ \vec{S}=(0,1,-1,0,0) \\
(2,-2,0)(2,-6,2)(1,3,2)(0,0,0)(0,0,0)
\end{array}$
\\ \hline

$\begin{array}{c}
\exp(-\phi+\vec{m}\cdot\vec{X}^-) \\ 
\vec{m}=(0,2,1,1,1) \\
(0,0,0)(2,2,0)(1,1,0)(1,1,0)(1,1,0)
\end{array}$
&
$\begin{array}{c}
(\partial X^+_2 + i\partial H_2)
\exp(\vec{m}\cdot\vec{X}^-+\vec{n}\cdot\vec{X}^++i\vec{S}\cdot\vec{H}+i\vec{\Lambda}\cdot\vec{\Phi}) \\
\vec{m}=(0,-2,2,0,0),\ \vec{n}=\left(0,\dfrac{4}{5},-\dfrac{1}{5},\dfrac{1}{5},\dfrac{1}{5}\right),\ \vec{S}=(0,1,-1,0,0) \\
(0,0,0)(2,-6,2)(1,3,2)(1,-1,0)(1,-1,0)
\end{array}$
\\ \hline

$\begin{array}{c}
\exp(-\phi+\vec{m}\cdot\vec{X}^-) \\ 
\vec{m}=(2,3,0,0,0) \\
(2,2,0)(3,3,0)(0,0,0)(0,0,0)(0,0,0)
\end{array}$
&
$\begin{array}{c}
\exp(\vec{m}\cdot\vec{X}^-+\vec{n}\cdot\vec{X}^++i\vec{S}\cdot\vec{H}+i\vec{\Lambda}\cdot\vec{\Phi}) \\
\vec{m}=(0,-1,1,0,0),\ \vec{n}=\left(\dfrac{2}{5},\dfrac{4}{5},-\dfrac{1}{5},0,0\right),\ \vec{S}=(0,1,-1,0,0) \\
(2,-2,0)(3,-5,2)(0,2,2)(0,0,0)(0,0,0)
\end{array}$
\\ \hline

$\begin{array}{c}
\exp(-\phi+\vec{m}\cdot\vec{X}^-) \\ 
\vec{m}=(0,3,0,1,1) \\
(0,0,0)(3,3,0)(0,0,0)(1,1,0)(1,1,0)
\end{array}$
&
$\begin{array}{c}
\exp(\vec{m}\cdot\vec{X}^-+\vec{n}\cdot\vec{X}^++i\vec{S}\cdot\vec{H}+i\vec{\Lambda}\cdot\vec{\Phi}) \\
\vec{m}=(0,-1,1,0,0),\ \vec{n}=\left(0,\dfrac{4}{5},-\dfrac{1}{5},\dfrac{1}{5},\dfrac{1}{5}\right),\ \vec{S}=(0,1,-1,0,0) \\
(0,0,0)(3,-5,2)(0,2,2)(1,-1,0)(1,-1,0)
\end{array}$
\\ \hline

\end{tabular}
}
\end{table}

\begin{table}[H]
\centering
{\small
\renewcommand{\arraystretch}{1.6}
\setlength{\tabcolsep}{8pt}

\begin{tabular}{|c|c|}
\hline
\textbf{Left-moving part ($V^L$)} & \textbf{Right-moving part ($V^R$)} \\ \hline\hline

$\begin{array}{c}
\exp(-\phi+\vec{m}\cdot\vec{X}^-) \\ 
\vec{m}=(2,0,0,0,3) \\
(2,2,0)(0,0,0)(0,0,0)(0,0,0)(3,3,0)
\end{array}$
&
$\begin{array}{c}
\exp(\vec{m}\cdot\vec{X}^-+\vec{n}\cdot\vec{X}^++i\vec{S}\cdot\vec{H}+i\vec{\Lambda}\cdot\vec{\Phi}) \\
\vec{m}=(0,0,0,1,-1),\ \vec{n}=\left(\dfrac{2}{5},0,0,-\dfrac{1}{5},\dfrac{4}{5}\right),\ \vec{S}=(0,0,0,-1,1) \\
(2,-2,0)(0,0,0)(0,0,0)(0,2,2)(3,-5,2)
\end{array}$
\\ \hline

$\begin{array}{c}
\exp(-\phi+\vec{m}\cdot\vec{X}^-) \\ 
\vec{m}=(0,1,1,0,3) \\
(0,0,0)(1,1,0)(1,1,0)(0,0,0)(3,3,0)
\end{array}$
&
$\begin{array}{c}
\exp(\vec{m}\cdot\vec{X}^-+\vec{n}\cdot\vec{X}^++i\vec{S}\cdot\vec{H}+i\vec{\Lambda}\cdot\vec{\Phi}) \\
\vec{m}=(0,0,0,1,-1),\ \vec{n}=\left(0,\dfrac{1}{5},\dfrac{1}{5},-\dfrac{1}{5},\dfrac{4}{5}\right),\ \vec{S}=(0,0,0,-1,1) \\
(0,0,0)(1,-1,0)(1,-1,0)(0,2,2)(3,-5,2)
\end{array}$
\\ \hline

$\begin{array}{c}
\exp(-\phi+\vec{m}\cdot\vec{X}^-) \\ 
\vec{m}=(2,0,0,1,2) \\
(2,2,0)(0,0,0)(0,0,0)(1,1,0)(2,2,0)
\end{array}$
&
$\begin{array}{c}
(\partial X^+_5 + i\partial H_5)
\exp(\vec{m}\cdot\vec{X}^-+\vec{n}\cdot\vec{X}^++i\vec{S}\cdot\vec{H}+i\vec{\Lambda}\cdot\vec{\Phi}) \\
\vec{m}=(0,0,0,2,-2),\ \vec{n}=\left(\dfrac{2}{5},0,0,-\dfrac{1}{5},\dfrac{4}{5}\right),\ \vec{S}=(0,0,0,-1,1) \\
(2,-2,0)(0,0,0)(0,0,0)(1,3,2)(2,-6,2)
\end{array}$
\\ \hline

$\begin{array}{c}
\exp(-\phi+\vec{m}\cdot\vec{X}^-) \\ 
\vec{m}=(0,1,1,1,2) \\
(0,0,0)(1,1,0)(1,1,0)(1,1,0)(2,2,0)
\end{array}$
&
$\begin{array}{c}
(\partial X^+_5 + i\partial H_5)
\exp(\vec{m}\cdot\vec{X}^-+\vec{n}\cdot\vec{X}^++i\vec{S}\cdot\vec{H}+i\vec{\Lambda}\cdot\vec{\Phi}) \\
\vec{m}=(0,0,0,2,-2),\ \vec{n}=\left(0,\dfrac{1}{5},\dfrac{1}{5},-\dfrac{1}{5},\dfrac{4}{5}\right),\ \vec{S}=(0,0,0,-1,1) \\
(0,0,0)(1,-1,0)(1,-1,0)(1,3,2)(2,-6,2)
\end{array}$
\\ \hline

$\begin{array}{c}
\exp(-\phi+\vec{m}\cdot\vec{X}^-) \\ 
\vec{m}=(2,0,0,2,1) \\
(2,2,0)(0,0,0)(0,0,0)(2,2,0)(1,1,0)
\end{array}$
&
$\begin{array}{c}
(\partial X^+_4 + i\partial H_4)
\exp(\vec{m}\cdot\vec{X}^-+\vec{n}\cdot\vec{X}^++i\vec{S}\cdot\vec{H}+i\vec{\Lambda}\cdot\vec{\Phi}) \\
\vec{m}=(0,0,0,-2,2),\ \vec{n}=\left(\dfrac{2}{5},0,0,\dfrac{4}{5},-\dfrac{1}{5}\right),\ \vec{S}=(0,0,0,1,-1) \\
(2,-2,0)(0,0,0)(0,0,0)(2,-6,2)(1,3,2)
\end{array}$
\\ \hline

$\begin{array}{c}
\exp(-\phi+\vec{m}\cdot\vec{X}^-) \\ 
\vec{m}=(0,1,1,2,1) \\
(0,0,0)(1,1,0)(1,1,0)(2,2,0)(1,1,0)
\end{array}$
&
$\begin{array}{c}
(\partial X^+_4 + i\partial H_4)
\exp(\vec{m}\cdot\vec{X}^-+\vec{n}\cdot\vec{X}^++i\vec{S}\cdot\vec{H}+i\vec{\Lambda}\cdot\vec{\Phi}) \\
\vec{m}=(0,0,0,-2,2),\ \vec{n}=\left(0,\dfrac{1}{5},\dfrac{1}{5},\dfrac{4}{5},-\dfrac{1}{5}\right),\ \vec{S}=(0,0,0,1,-1) \\
(0,0,0)(1,-1,0)(1,-1,0)(2,-6,2)(1,3,2)
\end{array}$
\\ \hline

$\begin{array}{c}
\exp(-\phi+\vec{m}\cdot\vec{X}^-) \\ 
\vec{m}=(2,0,0,3,0) \\
(2,2,0)(0,0,0)(0,0,0)(3,3,0)(0,0,0)
\end{array}$
&
$\begin{array}{c}
\exp(\vec{m}\cdot\vec{X}^-+\vec{n}\cdot\vec{X}^++i\vec{S}\cdot\vec{H}+i\vec{\Lambda}\cdot\vec{\Phi}) \\
\vec{m}=(0,0,0,-1,1),\ \vec{n}=\left(\dfrac{2}{5},0,0,\dfrac{4}{5},-\dfrac{1}{5}\right),\ \vec{S}=(0,0,0,1,-1) \\
(2,-2,0)(0,0,0)(0,0,0)(3,-5,2)(0,2,2)
\end{array}$
\\ \hline

$\begin{array}{c}
\exp(-\phi+\vec{m}\cdot\vec{X}^-) \\ 
\vec{m}=(0,1,1,3,0) \\
(0,0,0)(1,1,0)(1,1,0)(3,3,0)(0,0,0)
\end{array}$
&
$\begin{array}{c}
\exp(\vec{m}\cdot\vec{X}^-+\vec{n}\cdot\vec{X}^++i\vec{S}\cdot\vec{H}+i\vec{\Lambda}\cdot\vec{\Phi}) \\
\vec{m}=(0,0,0,-1,1),\ \vec{n}=\left(0,\dfrac{1}{5},\dfrac{1}{5},\dfrac{4}{5},-\dfrac{1}{5}\right),\ \vec{S}=(0,0,0,1,-1) \\
(0,0,0)(1,-1,0)(1,-1,0)(3,-5,2)(0,2,2)
\end{array}$
\\ \hline

\end{tabular}
}
\end{table}

\newpage
The first type of singlets (53 vertices) has the form
\begin{equation}
    \exp(-\phi+\vec{m}\vec{X}^-)(z)\otimes \exp(\vec{m}\vec{X}^--iH_j)(\bar{z}),\,\,\,m_j\neq0,
\end{equation}
where the vectors $\vec{m}$ correspond to deformations of $W_0$, invariant under $G$:
\begin{equation}
    \begin{array}{l}
        \vec{m}= (3,1,0,0,1),(3,0,1,0,1),(3,1,0,1,0),(3,0,1,1,0),\\
        (1,2,0,0,2),(1,0,2,0,2),(1,0,2,2,0),(1,2.0,2,0),\\
        (1,0,2,1,1),(1,1,1,0,2),(1,1,1,2,0),(1,2,0,1,1),(1,1,1,1,1),\\
        (0,0,0,2,3),(0,0,0,3,2),(0,3,2,0,0),(0,2,3,0,0).
    \end{array}
\end{equation}
These singlets are obtained from the $27$-multiplets by acting on the right-moving part with the operators $G^{-}_{i,-1/2}$ and replacing $\vec{\Lambda}$ by zero.

The second type of singlets (17 vertices corresponding to the mirror deformations):

\begin{table}[H]
\centering
{\small
\renewcommand{\arraystretch}{1.6}
\setlength{\tabcolsep}{8pt}

\begin{tabular}{|c|c|}
\hline
\textbf{Left-moving part ($V^L$)} & \textbf{Right-moving part ($V^R$)} \\ \hline\hline

$\begin{array}{c}
\exp(-\phi+\vec{m}\cdot\vec{X}^-) \\ 
\vec{m}=(1,0,0,2,2) \\
(1,1,0)(0,0,0)(0,0,0)(2,2,0)(2,2,0)
\end{array}$
&
$\begin{array}{c}
\exp(\vec{n}\cdot\vec{X}^++iH_j) \\ 
\vec{n}=\left(\dfrac{1}{5},0,0, \dfrac{2}{5},\dfrac{2}{5}\right), \ j \in \{1,4,5\} \\
(1,-1,2\delta_{1j})(0,0,0)(0,0,0)(2,-2,2\delta_{4j})(2,-2,2\delta_{5j})
\end{array}$
\\ \hline

$\begin{array}{c}
\exp(-\phi+\vec{m}\cdot\vec{X}^-) \\ 
\vec{m}=(1,1,1,1,1) \\
(1,1,0)(1,1,0)(1,1,0)(1,1,0)(1,1,0)
\end{array}$
&
$\begin{array}{c}
\exp(\vec{n}\cdot\vec{X}^++iH_j) \\ 
\vec{n}=\left(\dfrac{1}{5},\dfrac{1}{5},\dfrac{1}{5},\dfrac{1}{5},\dfrac{1}{5}\right), \ j \in \{1,2,3,4,5\} \\
(1,-1,2\delta_{1j})(1,-1,2\delta_{2j})(1,-1,2\delta_{3j})(1,-1,2\delta_{4j})(1,-1,2\delta_{5j})
\end{array}$
\\ \hline

$\begin{array}{c}
\exp(-\phi+\vec{m}\cdot\vec{X}^-) \\ 
\vec{m}=(1,2,2,0,0) \\
(1,1,0)(2,2,0)(2,2,0)(0,0,0)(0,0,0)
\end{array}$
&
$\begin{array}{c}
\exp(\vec{n}\cdot\vec{X}^++iH_j) \\ 
\vec{n}=\left(\dfrac{1}{5},\dfrac{2}{5},\dfrac{2}{5},0,0\right), \ j \in \{1,2,3\} \\
(1,-1,2\delta_{1j})(2,-2,2\delta_{2j})(2,-2,2\delta_{3j})(0,0,0)(0,0,0)
\end{array}$
\\ \hline

$\begin{array}{c}
\exp(-\phi+\vec{m}\cdot\vec{X}^-) \\ 
\vec{m}=(3,0,0,1,1) \\
(3,3,0)(0,0,0)(0,0,0)(1,1,0)(1,1,0)
\end{array}$
&
$\begin{array}{c}
\exp(\vec{n}\cdot\vec{X}^++iH_j) \\ 
\vec{n}=\left(\dfrac{3}{5},0,0,\dfrac{1}{5},\dfrac{1}{5}\right), \ j \in \{1,4,5\} \\
(3,-3,2\delta_{1j})(0,0,0)(0,0,0)(1,-1,2\delta_{4j})(1,-1,2\delta_{5j})
\end{array}$
\\ \hline

$\begin{array}{c}
\exp(-\phi+\vec{m}\cdot\vec{X}^-) \\ 
\vec{m}=(3,1,1,0,0) \\
(3,3,0)(1,1,0)(1,1,0)(0,0,0)(0,0,0)
\end{array}$
&
$\begin{array}{c}
\exp(\vec{n}\cdot\vec{X}^++iH_j) \\ 
\vec{n}=\left(\dfrac{3}{5},\dfrac{1}{5},\dfrac{1}{5},0,0\right), \ j \in \{1,2,3\} \\
(3,-3,2\delta_{1j})(1,-1,2\delta_{2j})(1,-1,2\delta_{3j})(0,0,0)(0,0,0)
\end{array}$
\\ \hline

\end{tabular}
}
\end{table}

The remaining 164 singlet vertices have a more general structure.

Here, we present several of them as examples:

\begin{table}[H]
\centering
{\small
\renewcommand{\arraystretch}{1.6}
\setlength{\tabcolsep}{8pt}

\begin{tabular}{|c|c|}
\hline
\textbf{Left-moving part ($V^L$)} & \textbf{Right-moving part ($V^R$)} \\ \hline\hline

$\begin{array}{c}
\exp(-\phi+\vec{m}\cdot\vec{X}^-) \\
\vec{m}=(1,1,1,2,0) \\
(1,1,0)(1,1,0)(1,1,0)(2,2,0)(0,0,0)
\end{array}$
&
$\begin{array}{c}
\exp(\vec{m}\cdot\vec{X}^-+\vec{n}\cdot\vec{X}^+ + i\vec{S}\cdot\vec{H}) \\
\vec{m}=(0,0,0,1,-1),\ \vec{n}=\left(\dfrac{1}{5},\dfrac{1}{5},\dfrac{1}{5},\dfrac{1}{5},\dfrac{1}{5}\right),\ \vec{S}=(0,0,0,0,1) \\
(1,-1,0)(1,-1,0)(1,-1,0)(2,0,0)(0,-2,2)
\end{array}$
\\ \hline

$\begin{array}{c}
\exp(-\phi+\vec{m}\cdot\vec{X}^-) \\
\vec{m}=(0,2,1,1,1) \\
(0,0,0)(2,2,0)(1,1,0)(1,1,0)(1,1,0)
\end{array}$
&
$\begin{array}{c}
(\partial X_2^- - i\partial H_2)
\exp(\vec{m}\cdot\vec{X}^-+\vec{n}\cdot\vec{X}^+ + i\vec{S}\cdot\vec{H}) \\
\vec{m}=(0,3,2,0,0),\ \vec{n}=\left(0,-\dfrac{1}{5},-\dfrac{1}{5},\dfrac{1}{5},\dfrac{1}{5}\right),\ \vec{S}=(0,0,-1,0,0) \\
(0,0,0)(2,4,0)(1,3,2)(1,-1,0)(1,-1,0)
\end{array}$
\\ \hline

$\begin{array}{c}
\exp(-\phi+\vec{m}\cdot\vec{X}^-) \\
\vec{m}=(2,2,0,0,1) \\
(2,2,0)(2,2,0)(0,0,0)(0,0,0)(1,1,0)
\end{array}$
&
$\begin{array}{c}
(\partial X_1^+ + i\partial H_1)
(\partial X_3^- - i\partial H_3)
\exp(\vec{m}\cdot\vec{X}^-+\vec{n}\cdot\vec{X}^+ + i\vec{S}\cdot\vec{H}) \\
\vec{m}=(-2,-1,2,0,1),\ \vec{n}=\left(\dfrac{4}{5},\dfrac{3}{5},-\dfrac{2}{5},0,0\right),\ \vec{S}=(1,1,-1,0,0) \\
(2,-6,2)(2,-4,2)(0,4,2)(0,0,0)(1,1,0)
\end{array}$
\\ \hline

$\begin{array}{c}
\exp(-\phi+\vec{m}\cdot\vec{X}^-) \\
\vec{m}=(2,1,1,1,0) \\
(2,2,0)(1,1,0)(1,1,0)(1,1,0)(0,0,0)
\end{array}$
&
$\begin{array}{c}
(\partial X_2^+ + i\partial H_2)
(\partial X_3^+ + i\partial H_3)
\exp(\vec{m}\cdot\vec{X}^-+\vec{n}\cdot\vec{X}^+ + i\vec{S}\cdot\vec{H}) \\
\vec{m}=(3,-2,-2,1,0),\ \vec{n}=\left(-\dfrac{1}{5},\dfrac{3}{5},\dfrac{3}{5},0,0\right),\ \vec{S}=(-1,1,1,0,0) \\
(2,4,2)(1,-5,2)(1,-5,2)(1,1,0)(0,0,0)
\end{array}$
\\ \hline

\end{tabular}
}
\end{table}


\end{document}